\documentclass[a4paper,12pt]{article}

\usepackage[pages=all, color=black, position={current page.south}, placement=bottom, scale=1, opacity=1, vshift=5mm]{background}
\SetBgContents{
	\tt This work is shared under a \href{https://creativecommons.org/licenses/by-sa/4.0/}{CC BY-SA 4.0 license} unless otherwise noted
}      % copyright

\usepackage[margin=1in]{geometry} % full-width

\usepackage{amsmath}
\usepackage{amsthm}
\usepackage{mathtools}
\usepackage{amssymb}
\usepackage{bbm}
\usepackage{rotating}
\usepackage{pdflscape}
\usepackage{array,longtable}
\usepackage{xltabular}
\usepackage{soul}
\usepackage{textcomp} % celsius symbol
\usepackage{mathcomp} % celsius symbol
\usepackage{makecell}
\usepackage{multirow,bigdelim}
\usepackage{array, makecell}
\usepackage{rotfloat}
\usepackage{caption}

\newtheorem{thm}{Theorem}

\usepackage[utf8]{inputenc}
\usepackage{hyperref}
\hypersetup{
	unicode,
	pdfauthor={Author One, Author Two, Author Three},
	pdftitle={A simple article template},
	pdfsubject={A simple article template},
	pdfkeywords={article, template, simple},
	pdfproducer={LaTeX},
	pdfcreator={pdflatex}
}

\usepackage[sort&compress,numbers,square]{natbib}
\newcommand{\argmin}{\arg\!\min}

\theoremstyle{plain}

\theoremstyle{definition}

\usepackage{graphicx, color}
\graphicspath{{fig/}}

\newcommand{\E}{\mathrm{E}}

\newcommand{\Var}{\mathrm{Var}}

\usepackage{algorithm, algpseudocode} % use algorithm and algorithmicx for typesetting algorithms
\usepackage{mathrsfs} % for \mathscr command

\usepackage{lipsum}

\newcolumntype{L}[1]{>{\raggedright\let\newline\\\arraybackslash\hspace{0pt}}m{#1}}
\newcolumntype{C}[1]{>{\centering\let\newline\\\arraybackslash\hspace{0pt}}m{#1}}
\newcolumntype{R}[1]{>{\raggedleft\let\newline\\\arraybackslash\hspace{0pt}}m{#1}}

\title{Detection of Multiple Inf\hspace{+0.05em}luential Observations on Model Selection}
\author{Dongliang Zhang$^1$ \and Masoud Asgharian$^2$ \and Martin A.~Lindquist$^1$}

\date{
	$^1$Department of Biostatistics, Johns Hopkins University, Baltimore, Maryland, U.S.A. \\  %\texttt{Emails: dzhang69@jhu.edu; masoud.asgharian2@mcgill.ca; mlindqui@jhsph.edu}\\%
	$^2$Department of Mathematics and Statistics, McGill University, Montr\'{e}al, Qu\'{e}bec, Canada \\ %\texttt{auth3@inst2.edu}\\[2ex]%
}

\begin{document}

\maketitle
	
\begin{abstract}
Outlying observations are frequently encountered across a wide spectrum of scientific domains, posing notable challenges to the generalizability of statistical models and the reproducibility of downstream analysis. They are identified through inf\hspace{+0.05em}luential diagnostics, which aim to capture observations that unduly bias model estimation. To date, methods for identifying observations that inf\hspace{+0.05em}luence the selection of a stochastically chosen submodel have been underdeveloped, especially in the high-dimensional setting where the number of predictors $p$ exceeds the sample size $n$. Recently we proposed an improved diagnostic measure to handle this setting. However, its distributional properties and approximations have not yet been explored. To address this shortcoming, we revisit the notion of exchangeability to determine the exact asymptotic distribution of our assessment measure. This foundation enables the introduction of theoretically supported parametric and nonparametric approaches for distributional approximation and derivation of thresholds for outlier identification. The resulting framework is further extended to logistic regression models and evaluated by comprehensive simulation studies comparing the performance of various detection methods. Finally, the framework is applied to data from a task-based fMRI study of thermal pain, with the goal of identifying outliers that distort the formulation of the statistical model using functional brain activity to predict physical pain ratings. Both linear and logistic models are used to demonstrate the benefits of detection and compare the performance of dif{}ferent detection procedures. In particular, we identify two inf\hspace{+0.05em}luential observations that were not detected in prior studies

\noindent\textbf{Keywords:} clustering analysis; fMRI; high-dimensional diagnosis; inf\hspace{+0.05em}luential point detection; logistic regression models; variable selection
\end{abstract}

\newpage 

\tableofcontents

\newpage 
	
\section{Introduction}
\label{sec:introduction}

%In the era of ``Big Data'', technological advancements have enabled unprecedented capability to generate, process, and analyze large-scale datasets. Yet, their sheer volume and complex structure do not necessarily confer the characteristics required for accurate and reproducible inference. One major hindrance is data heterogeneity in the form of outlying measurements. 

In the era of ``Big Data'', technological advancements have enabled unprecedented capacity to generate and analyze large-scale datasets. Yet, their volume and complexity do not guarantee accurate or reproducible inference, as outlier-driven heterogeneity remains a major challenge.

%Consider a neuroimaging pain study leveraging as a motivating example for the need to identify and address anomalous data points

\textcolor{black}{Consider a neuroimaging pain study that motivates the need to detect and address anomalous data points. Specifically, in predicting physical pain from task-based functional magnetic resonance imaging (fMRI) data, the presence of multiple outlying measurements can compromise model generalizability. They may arise from inter-individual variability in pain-related brain activation, misalignment due to heterogeneous functional topology across subjects, and motion artifacts during intense stimulation. These factors can together produce multiple outliers that distort model fitting, undermining generalizability and the reliability of downstream statistical analysis. Hence, inf\hspace{+0.05em}luential diagnosis should be performed to identify subset of data unduly inf\hspace{+0.05em}luencing model estimates. Since inf\hspace{+0.05em}luential points are prevalent across disciplines, these detection methods should be broadly applicable beyond neuroimaging to diverse data types and scientific fields concerned with model generalizability and inferential validity}. 

In the low-dimensional setting where the number of predictors $p$ is less than the sample size $n$, inf\hspace{+0.05em}luence on deterministic submodels has been well studied. Yet, in practice, submodels are usually decided in a data-driven manner. When $p$ exceeds $n$, high-dimensional model selectors such as LASSO \citep{lasso}, scaled LASSO (SLASSO) \citep{sun2012}, SCAD \citep{fan2001}, and MCP \citep{zhang2010} are widely used to derive essential information.

To date, identifying observations that distort stochastic selection of submodels has received limited attention. Indeed, multiple inf\hspace{+0.05em}luential observations can markedly increase the probability of selecting an incorrect submodel when they are only partially or incorrectly excluded (Web Figure S1), underscoring the crucial necessity for their comprehensive identification.

To detect them, \cite{zhao2013} proposed the high-dimensional inf\hspace{+0.05em}luence measure (HIM) to capture the inf\hspace{+0.05em}luence of a single observation on the correlation between the response and predictors. This was extended by \cite{zhao2019} to detect multiple outliers, leading to the multiple inf\hspace{+0.05em}luential points (MIP) detection method. \cite{bala2019} subsequently introduced the DF(LASSO) metric to assess variation in models selected by LASSO. Building on this line of work, \cite{zhang2023} established the central limit theorem (CLT) for the generalized dif{}ference in model selection (GDF) metric, applicable to general model selectors. A consistent high-dimensional clustering method was then integrated with the GDF measure, resulting in the clustering-based MIP (ClusMIP) algorithm to detect multiple outliers. 

However, the works of \cite{bala2019} and \cite{zhang2023} are limited in two ways: (\textit{i}) the distributional properties of the DF(LASSO) and GDF metrics remain unexplored, and (\textit{ii}) no principled methods exist for accurately approximating their distributions.

%\citep{skellam2018}: BB
%\citep{consul1973}: GP
%\footnote{Available to be downloaded from https://github.com/Dongliang-JHU/ClusMIP.}

This manuscript aims to consolidate and expand the findings of our previous work \citep{zhang2023} by \textcolor{black}{advancing study across theoretical, methodological, and application domains}. 

\textcolor{black}{On the theoretical front, while the exchangeability property and the CLT of the DF(LASSO) and GDF measures were established previously, the current work further justifies that these measures are asymptotically as $p \rightarrow \infty$ mixtures of binomial random variables. This result leverages the characterization of sums of arbitrarily dependent sequence of events by unique exchangeable counterparts \citep{galambos1978} and the classical de Finetti's Representation Theorem \citep{finetti31}. Based on these results, the theoretical basis for approximating their distributions via parametric approach is established.}

\textcolor{black}{On the methodological front, the earlier work relied only on the CLT-dependent approximation under $\min(n,p) \rightarrow \infty$, limiting its practical utility. Here, we broaden the suite of approximation options by introducing both parametric and nonparametric approaches for deriving diagnostic thresholds using the mid-quantiles. Specifically, the parametric option, which includes the Conway-Maxwell-Binomial, Conway-Maxwell-Poisson, beta-binomial, generalized Poisson, and mixtures of binomial and Poisson distributions, of{}fers valuable insight into the latent dispersion or association structure of the diagnostic metrics, enhancing interpretability. The nonparametric approach, which solely requires a large $n$, provides a robust alternative, encompassing three bootstrap schemes \citep{muliere1992, jentsch2016} specifically tailored for discrete and exchangeable data. These approximation strategies are incorporated into the ClusMIP procedure, yielding a revised version that also extends to logistic regression models, which is implemented in the newly developed $\tt{R}$ package $\tt{ClusMIP}$.} 

\textcolor{black}{On the application front, by applying both linear and logistic regression models to the same fMRI dataset from the thermal pain study \citep{lindquist2017} with the newly updated approximation options, the present work identifies additional inf\hspace{+0.05em}luential observations and thereby extends the conclusions of the previous research}.  

%Extension to the logistic regression model is discussed in Section~\ref{sec:multiple}.

\textcolor{black}{The article is organized as follows: the existing literature is surveyed in Section~\ref{sec:existing}}. \cite{zhang2023} is then revised and improved in Section~\ref{sec:revival}. This is followed by introducing the parametric and nonparametric approximations with their choice guidelines from Sections~\ref{sec:parametric} to \ref{sec:guidelines}, respectively. Extending our work to the logistic regression model is given in Section~\ref{sec:multiple}. Ssections~\ref{sec:simulation} and \ref{sec:real} include simulations and the outlier detection analysis of an fMRI dataset collected to predict physical pain \citep{lindquist2017}. Conclusions are given in Section~\ref{sec:conclusions}.

\section{Review of Existing Literature}
\label{sec:existing}

For $i=1,...,n$, consider the linear regression model with \emph{i.i.d.}\ errors $\sigma^2 > 0$:
\begin{equation}
\text{Y}_{i}=\textbf{X}_{i}^\top \boldsymbol{\beta} + \epsilon_i, \ \epsilon_{i} \sim N(0,\sigma^2), \label{linearmodel}
\end{equation} 
where $\textbf{Y}=(\text{Y}_{1},...,\text{Y}_{n})^{\top} \in \mathbb{R}^{n}$ is the response vector, $\textbf{X}_{i}=(\text{X}_{i1},...,\text{X}_{ip})^{\top} \in \mathbb{R}^{p}$ is the $i^{\text{th}}$ row of the design matrix $\textbf{X}$, and $\boldsymbol{\beta}=(\beta_1,...,\beta_p)^{\top} \in \mathbb{R}^{p}$ is the regression coef{}ficient. We next summarize existing methods, while comprehensive details are given in Web Appendix B. 

\noindent \emph{\bf HIM and MIP}. Under \eqref{linearmodel}, \cite{zhao2013} introduced the high-dimensional inf\hspace{+0.05em}luence measure (HIM) to capture the ef{}fect of a single observation on the correlation between the response and predictors. Under suitable conditions, the scaled HIM measure asymptotically follows the chi-square distribution, providing a diagnostic threshold. Building on this, \cite{zhao2019} incorporated the so-called random-group-deletion (RGD) strategy into the HIM framework, yielding the multiple inf\hspace{+0.05em}luential points (MIP) procedure. Such a scheme mitigates the swamping and masking ef{}fects in multiple outlier detection through repeatedly attaching a candidate observation to randomly drawn subsets of the remaining data. When the number of subsets is suf{}ficiently large, this method ensures that, with high probability, each influential point appears as the sole contamination in at least one subset, enabling its accurate detection.  

\noindent {\bf DF(LASSO)}. \cite{bala2019} proposed the DF(LASSO) metric to assess variation in submodels selected by LASSO, defined as $
\delta_i = \sum_{j=1}^{p} \big| \mathbbm{1}(\widehat{\beta}_{\text{LASSO},j}=0) - \mathbbm{1}(\widehat{\beta}^{(i)}_{\text{LASSO},j}=0) \big|$. 
Here, $\widehat{\beta}_{\text{LASSO},j}$ and $\widehat{\beta}^{(i)}_{\text{LASSO},j}$ denote the $j^\text{th}$ LASSO estimates from the complete and reduced datasets removing the $i^\text{th}$ point, and $\mathbbm{1}(\cdot)$ is the indicator function. The CLT of $\delta_i$ was shown under the low-dimensional setting $n<p$, of{}fering the basis for deriving diagnostic threshold.

\section{Methodology}
\label{sec:methodology}

\subsection{Revival of Exchangeability}
\label{sec:revival}

\cite{zhang2023} relaxed the restriction on model selectors used \textcolor{black}{in the DF(LASSO)} metric and introduced the following generalized dif{}ference in model selection (GDF) metric:
\begin{equation}
\tau_i = \sum_{j=1}^{p}  \lvert \mathbbm{1}(\widehat{\beta}_{j}=0) - \mathbbm{1}( \widehat{\beta}^{(i)}_{j}=0) \big \rvert, \ i=1,...,n, \label{gdf}
\end{equation}
where $\widehat{\beta}_{j}$ and $\widehat{\beta}^{(i)}_{j}$ denote the $j^\text{th}$ elements of the sparse coefficient estimates obtained from the full and reduced datasets excluding the $i^\text{th}$ observation, respectively. To approximate the distribution of $\tau_i$, \cite{zhang2023} derived the CLT for $\tau_i$ as $\text{min}(n,p) \rightarrow \infty$ [\emph{ibid.}, Theorem 2] by characterizing $\tau_i=\sum_{j=1}^{p} \xi_{ij}$, where $\xi_{[p]}=(\xi_{i1},\ldots,\xi_{ip})$ forms an exchangeable Bernoulli sequence \citep{galambos1978} (suppressing the subscript $i$), and applying CLT for such exchangeable sequence \citep{klass1987}. The sample mean and variance of $\tau_{[n]}$ were further justified to consistently estimate $\mathbb{E}(\tau_i)$ and $\Var(\tau_i)$ conditionally, supported via the preservation of exchangeability \citep{commenges2003}, uniform consistency of the empirical distribution and strong law of large numbers for an exchangeable sequence. Here, $a_{[n]}$ denotes an arbitrary sequence $a_1,\ldots,a_n$. Yet, \cite{zhang2023} did not include a systematic study of (\emph{i}) the distributional properties of $\tau_i$ and (\emph{ii}) the associated alternative approximations.

To examine the distribution of $\tau_i$, the de Finetti's Representation Theorem (DFRT) implies that as $p \rightarrow \infty$, $\xi_{[p]}$ is \emph{i.i.d.} conditioning on a latent variable $\Theta$ governed by the probability measure $\mu_{\infty}$, where $\Theta$ denotes the limiting relative frequency of successes and $\mu_{\infty}$ the limiting distribution of the empirical measure. Hence, asymptotically, $\xi_{[p]}$ is an unconditional mixture of infinite \emph{i.i.d.} Bernoulli sequences, and each $\tau_i$ is the sum of such a mixture. Since the sum of \emph{i.i.d.} binaries  follows a binomial law, the large-sample distribution of $\tau_i$ in \eqref{gdf} as $p \rightarrow \infty$ is stated in the theorem below, with its proof given in Web Appendix C. The same distributional result holds for the DF(LASSO) measure $\delta_i$, since it is a special case where LASSO is used.

\begin{thm}[]\label{thm:tau_asymptotic}
Let $\tau_i$ denote the GDF metric in \eqref{gdf} and $\delta_{i}$ be the DF(LASSO) measure defined above. Then, as $p \rightarrow \infty$, both $\tau_i$ and $\delta_i$ follow finite mixtures of binomial distributions.
\end{thm}

Having established the large-sample distributions of the DF(LASSO) and GDF measures, we next characterize the distributional properties of the de Finetti mixing measure $\mu_{\infty}$. Under an exchangeable sampling scheme from a basic P\'{o}lya urn model with two colored balls, $\mu_\infty$ follows a beta distribution. Extending such scheme to more than two colors yields a Dirichlet distribution. This connection is formalized in the works of \cite{johnson1932} and \cite{zabell1982}, who substantiated that for an exchangeable sequence of discrete random variables $\text{X}_{[n]}$, if the predictive probability $\mathbb{P}(\text{X}_{n+1}\mid \text{X}_{[n]})$ exhibits a linear form for all $n$, then $\mu_\infty$ is Dirichlet. This assumption, known as Johnson’s ``sufficientness'' postulate, traces philosophically back to Laplace's Rule of Succession for enumerative induction. These results collectively provide insight into the characterization of $\mu_{\infty}$. See \cite{wagenmakers2023} for a recent account.

On the other hand, both the DFRT and Theorem~\ref{thm:tau_asymptotic} depend on asymptotic arguments and thus do not hold for the finite exchangeable sequence of binaries $\xi_{[p]}$ \citep{diaconis1980}. Nevertheless, the finite-sample variant of the DFRT [\emph{ibid.}, Theorem 3] shows that when $p$ is large but finite, the distribution of the first $k$ elements of $\xi_{[p]}$ remains close to the corresponding de Finetti-type representation. This suggests that \textcolor{black}{binomial} mixtures provide a reasonable approximation to the finite-sample distribution of $\tau_i$ in \eqref{gdf} for large, finite $p$. 

%with non-negative pairwise covariance as $p \rightarrow \infty$ (Lemma~\ref{lemma:exchangeable_variance})

In summary, we have (\emph{i}) $\tau_{[n]}$ is exchangeable and thus marginally identically distributed, (\emph{ii}) each $\tau_i$ is identically distributed as the sum of $p$ exchangeable binaries in finite sample, and (\emph{iii}) each $\tau_i$ follows a mixture of binomial distributions as $p \rightarrow \infty$ (Theorem~\ref{thm:tau_asymptotic}). They form the basis for alternative approximations to $\tau_i$ to be presented in Sections~\ref{sec:parametric} and~\ref{sec:nonparametric}.

\subsection{Approximation via Parametric Approaches}
\label{sec:parametric}

We note that the $\tau_{[n]}$ are not independent. However, exchangeability ensures that the $\tau_i$'s are marginally identically distributed. Let $F_{\tau}$ denote their common distribution. To approximate $F_{\tau}$ parametrically, we first study the legitimacy of the maximum likelihood estimation (MLE) principle. Under general regularity conditions, Theorem~\ref{thm:kldistance} establishes the consistency of the minimizer of the Kullback–Leibler (K–L) divergence between $F_{\tau}$ and the postulated family of parametric distributions. The related  conditions and proof are provided in Web Appendix~D.

\begin{thm}[]\label{thm:kldistance}
Let $G_{\boldsymbol{\theta}}$ be a parametric distribution indexed by the $q$-dimensional parameter $\boldsymbol{\theta} \in \Theta \subseteq \mathbb{R}^q$, where $\Theta$ is a bounded open set, and $F_{\tau}$ be the true common distribution function of $\tau_i$'s. Suppose conditions (A.1) to (A.5) (see Web Appendix D) hold. Then $\widehat{\boldsymbol{\theta}}_n \xrightarrow{p} \boldsymbol{\theta}^*$ as $n \rightarrow \infty$, where $\widehat{\boldsymbol{\theta}}_n$ is the MLE estimator obtained upon fitting $G_{\boldsymbol{\theta}}$ to $\tau_{[n]}$ as if \textcolor{black}{$\tau_{[n]}$ was independent}, and 
$\boldsymbol{\theta}^* = \argmin_{\theta \in \Theta} \ D_{KL} \left(F_{\tau} \ || \ G_{\boldsymbol{\theta}} \right)$, the K-L divergence between $F_{\tau}$ and $G_{\boldsymbol{\theta}}$.
\end{thm}

Founded on the exchangeability of $\tau_{[n]}$, Theorem~\ref{thm:kldistance} advocates the use of MLE even when independence does not hold. The MLE is asymptotically tantamount to minimizing the K-L divergence between $\widehat{F}_n$ and $G_{\boldsymbol{\theta}}$, given by $\text{I}(\widehat{F}_n, G_{\boldsymbol{\theta}}) = \E_{\widehat{F}_n}\!\left[\log{\frac{d\widehat{F}_n}{dG_{\boldsymbol{\theta}}}}\right]$, where $\widehat{F}_n$ is the empirical distribution of $\tau_{[n]}$. Since $\widehat{F}_n \xrightarrow{\text{a.s.}} F_{\tau}$ as $n \rightarrow \infty$ \citep{athreya2016}, the estimator $\widehat{\boldsymbol{\theta}}_n$ consistently estimates $\boldsymbol{\theta}^*$, where $G_{\boldsymbol{\theta}^*}$ is the closest element to $F_{\tau}$ in terms of K-L divergence that can be achieved within the possibly misspecified family of parametric distributions $G_{\boldsymbol{\theta}}$.

%relative to the Poisson law

\textcolor{black}{In view of Theorem~\ref{thm:kldistance} and the established distributional properties, we adopt three guiding principles for selecting parametric candidates. First, each $\tau_i$ in \eqref{gdf} is the sum of $p$ Bernoulli variables with intractable dependence. Instead of estimating each Bernoulli component marginally, a more viable option is to directly model their holistic behavior, capturing the overall dependence through a parameter quantifying both its direction and magnitude. Second, since $\tau_{[n]}$ forms a sequence of count statistics ref\hspace{+0.05em}lecting discrepancies in selected submodels and asymptotically follows a binomial-type mixture, a Poisson-type approximation is reasonable. However, this mixture structure implies potential dif{}fering mean and variance across subpopulations, yielding dispersion. Indeed, an observation that biases the choice of one predictor may simultaneously af{}fect the selection of other variables. Such non-negative dependence may inflate the variance of the sum, resulting in over-dispersion (Web Figure S2). This motivates the use of an over-dispersed Poisson-type model. Lastly, Theorem~\ref{thm:tau_asymptotic} further supports parametric fitting approaches consistent with these distributional properties}. 

%Their detailed formulations, together with their correspondence to the above guidelines is given in Web Appendix E

\textcolor{black}{Following these guidelines, we consider six parametric families satisfying conditions (A.1)-(A.5) of Theorem~\ref{thm:tau_asymptotic}: the Conway-Maxwell-Binomial (CMB), Conway-Maxwell-Poisson (CMP), beta-binomial (BB), generalized Poisson (GP), mixtures of binomial (MB) and Poisson (MP) distributions. Formulation and adherence to these guidelines is given in Web Appendix E}. 

%For a parametric candidate $G_{\boldsymbol{\theta}}$, let the empirical and parametric counterparts of $Q_{\zeta,\text{mid}}$ be $\widehat{Q}_{\zeta,\text{mid}}$ and $\widehat{Q}_{\zeta,\text{mid},G_{\boldsymbol{\theta}}}$, respectively.
 
Due to the discreteness of $\tau_{[n]}$, the sample quantile of the parametric candidates is inconsistent for estimating the true quantile. To address this issue, we consider a variant known as mid-quantile, which smooths the discontinuities in discrete distributions by interpolating probability masses, yielding stable and asymptotically reliable thresholds. Let $Q_{\zeta,\text{mid}}$ and $\widehat{Q}_{\zeta,\text{mid}}$ be the $(100\times\zeta)^\text{th}$ population and sample mid-quantiles of $\tau_{[n]}$ respectively, $\zeta \in (0,1)$. It is shown \citep{ma2011} that $\sqrt{n}(\widehat{Q}_{\zeta,\text{mid}} - Q_{\zeta,\text{mid}})$ is asymptotically normal if $\zeta$ does not correspond to boundary values, forming the theoretical validity of $\widehat{Q}_{\zeta,\text{mid}}$. However, it may be restrictive in practice since outliers are always identified with $\widehat{Q}_{\zeta,\text{mid}}$, even for clean dataset. In view of that, we consider $\widehat{Q}_{\zeta,\text{mid},G_{\boldsymbol{\theta}}}$, the mid-quantile of the diagnostic metrics estimated based on the parametric candidate $G_{\boldsymbol{\theta}}$. Despite lacking an established convergence result for $\sqrt{n}(\widehat{Q}_{\zeta,\text{mid},G_{\boldsymbol{\theta}}}-Q_{\zeta,\text{mid}})$, its decomposition as $\sqrt{n} (\widehat{Q}_{\zeta,\text{mid},G_{\boldsymbol{\theta}}}-\widehat{Q}_{\zeta,\text{mid}}) + \sqrt{n} (\widehat{Q}_{\zeta,\text{mid}}-Q_{\zeta,\text{mid}})$ of{}fers valuable insights: while the second term vanishes asymptotically (see above), the first term remains reasonably controlled since $\widehat{Q}_{\zeta,\text{mid},G_{\boldsymbol{\theta}}}$ is obtained by fitting $G_{\boldsymbol{\theta}}$ to $\tau_{[n]}$. Thus, $\widehat{Q}_{\zeta,\text{mid},G_{\boldsymbol{\theta}}}$ provides an alternative threshold, where its practical ef{}fectiveness is shown by simulations.  

\subsection{Approximation via Nonparametric Approaches} 
\label{sec:nonparametric}

\textcolor{black}{We present three established bootstrap schemes from the literature, applied here solely for the purpose of obtaining thresholds for $\tau_{[n]}$, without any claim of methodological novelty}.

% as $n \rightarrow \infty$ 

\noindent \emph{\bf Method I}. \citep{muliere1992} applied bootstrap directly on $\tau_{[n]}$ to infer about its conditional average. By the DFRT and its related extension, there exists a random variable $\Gamma$ conditioning on which $\tau_{[n]}$ is \emph{i.i.d.} from an exponential family distribution, and $\Gamma$ is also the limiting law of $T_n=T_n(\tau_{[n]})=\sum_{i=1}^{n}\tau_i/n$ [\emph{ibid.}, Theorem 3.2]. Resampling $\tau_{[n]}$ to obtain $\tau^{*}_{[n]}$, the empirical distribution of $T_n^*=T_n(\tau^{*}_{[n]})$ reasonably approximates the distribution of $\Gamma \vert \tau_{[n]}$ [\emph{ibid.}, Theorem 3.4]. Moreover, it is further verified [\emph{ibid.}, Theorem 3.7] that $Z_n^* \coloneq \sqrt{n}(T_{n}^{*}-T_{n})/S_{n} \vert \tau_{[n]}$ is a standard normal as $n \rightarrow \infty$, where $S_n^2=\sum_{i=1}^n(\tau_i-T_n)^2/n$. Leveraging the CLT for exchangeable sequence, $Z_n \coloneq \sqrt{n}(T_n-\gamma)/S_n$ is also a standard normal as $n \rightarrow \infty$, given $\Gamma=\gamma$. \textcolor{black}{This produces the bootstrap consistency result: $\sup_{t} \vert P(Z_n^* \leq t \vert \tau_1,...,\tau_n) - P(Z_n \leq t) \vert\xrightarrow[n \rightarrow \infty]{} 0$, which guarantees the asymptotic coverage of the bootstrap percentile confidence interval (CI)}. In practice, its lower bound of{}fers an ef{}fective diagnostic threshold.

\noindent \emph{\bf Methods II and III}. \cite{jentsch2016} studied two $m$-out-of-$n$ bootstrap schemes, resampling $m$ observations with replacement from $\tau_{[n]}$ to consistently approximate the distributions of the sample quantile $\widehat{Q}_{\zeta}$ and true mid-quantile $Q_{\zeta,\text{mid}}$ by the bootstrap counterparts. Suppose that $m/n+1/m=o(1)$. Then, \textcolor{black}{the Kolmogorov-Smirnov distance between $\widehat{Q}_{\zeta,m}^*$ and $\widehat{Q}_{\zeta}$, $d_{KS}(\widehat{Q}_{\zeta,m}^*, \widehat{Q}_{\zeta})$, vanishes [\emph{ibid.}, Theorem 8] as $n \rightarrow \infty$, where $\widehat{Q}_{\zeta,m}^*$ is the bootstrap analogue of $\widehat{Q}_{\zeta}$. Such bootstrap consistency  leads to the confidence set (CS) based on $\widehat{Q}_{\zeta,m}^*$, of{}fering  guarantee on the asymptotic coverage of the true quantile $Q_{\zeta}$ [\emph{ibid.}, Theorem 11] as $P(Q_{\zeta} \in \text{CS}) \xrightarrow[n \rightarrow \infty]{} 1-\alpha$ for $\alpha \in (0,0.5)$}. Similarly, \textcolor{black}{$d_{KS}(\sqrt{m}(\widehat{Q}_{\zeta,\text{mid},m}^*-\widehat{Q}_{\zeta,\text{mid}}), \sqrt{n}(\widehat{Q}_{\zeta,\text{mid}}-Q_{\zeta,\text{mid}})) \xrightarrow[n \rightarrow \infty]{} 0$ [\emph{ibid.}, Corollary 18], where $\widehat{Q}_{\zeta,\text{mid}}$ is the sample mid-quantile and $\widehat{Q}_{\zeta,\text{mid},m}^*$ is its bootstrap analogue. Such bootstrap consistency gives rise to CI formed from $\widehat{Q}_{\zeta,\text{mid}}$ and $\widehat{Q}_{\zeta,\text{mid},m}^*$ with asymptotic coverage guarantee of the true mid-quantile $Q_{\zeta,\text{mid}}$ [\emph{ibid.}, Theorem 19] as $P(Q_{\zeta,\text{mid}} \in \text{CI}) \xrightarrow[n \rightarrow \infty]{} 1-\alpha$}. In practice, the averaged bootstrap estimates of sample- and mid-quantiles serve as ef{}fective working diagnostic thresholds as verified in simulations.

\subsection{Guideline on the Choices of Approximation Approaches} 
\label{sec:guidelines}

%\citep{zhang2023} 
%Detailed discussions are given in Web Appendix F
%as discussed in Section~\ref{sec:parametric},

\textcolor{black}{We consider four factors in choosing approximation methods: dimensionality, interpretability, robustness and computational cost. First, while the CLT-based approach requires both large $n$ and $p$, the parametric and nonparametric methods rely primarily on large $n$, with large $p$ further ensuring accurate parametric fitting (Theorem~\ref{thm:tau_asymptotic}). Thus, the nonparametric method is the least restrictive option on dimensionality. Moreover, the parametric approach enhances interpretability by revealing the latent structure of $\tau_{[n]}$ (Section~\ref{sec:parametric}). In contrast, the nonparametric alternative, which is distribution-free, is more robust to model misspecification but at the expense of higher computational cost depending on the number of bootstrap replicates. Hence, we recommend the nonparametric approach as the default choice for robustness, the parametric approach when interpretability or computational ef{}ficiency is prioritized, and the CLT-based method as a baseline benchmark. See Web Appendix F for detailed discussions}.

\subsection{Detection of Multiple Inf\hspace{+0.05em}luential Observations} 
\label{sec:multiple}

\cite{zhang2023} introduced the ClusMIP procedure based on the GDF metric \eqref{gdf}, denoted by ClusMIP(Selector) with a given model selector, for multiple outlier detection. Specifically, the data are first partitioned into potentially inf\hspace{+0.05em}luential ($\widehat{\text{S}}_{\text{inf\hspace{+0.05em}l}}$) and clean ($\widehat{\text{S}}_{\text{clean}}$) portions via a consistent high-dimensional clustering scheme. Each point $i \in \widehat{\text{S}}_{\text{inf\hspace{+0.05em}l}}$ is then assessed using the merged sample $\{i\} \cup \widehat{\text{S}}_{\text{clean}}$, where it would be the sole outlier if it is truly contaminated. Detection proceeds via single-point inference using CLT or the newly proposed parametric and nonparametric methods. \textcolor{black}{This extension broadens the approximation toolbox of the original ClusMIP, leading to a revised version (Web Appendix G) with nine additional approximation methods, thereby considerably enhancing its practical applicability. The updated version has been implemented in the R package \texttt{ClusMIP}, of{}fered on the first author’s GitHub page. To further facilitate the choice of dif{}ferent detection and approximation options, the key features of the existing methods and the revised ClusMIP procedure are summarized in Table~\ref{tab:method_comparison}}.  

\begin{table}[H]
\centering
 \addtolength{\leftskip}{-2cm}
    \addtolength{\rightskip}{-2cm}
    \fontsize{7.6}{9.1} \selectfont
    \caption{\centering Comparison of Existing and Proposed Detection Procedures.}
\begin{tabular}{C{2cm} C{1.9cm} C{3.4cm} C{3.4cm} C{3.6cm} C{3.2cm}} \hline
& \textbf{Metrics} & \textbf{HIM} & \textbf{MIP} & \textbf{DF(LASSO)} & \textbf{\makecell{ClusMIP}} \\ \hline \hline

\multirow{3}{*}{\textbf{Design}} & \textbf{Purpose} & model-free, correlation-based single outlier detection & model-free, correlation-based multiple outlier detection & detection of a single outlier on model selected by LASSO & detection of multiple outliers on models selected by generic model selectors \\ \cline{2-6}

& \textbf{Multiple Outliers} & No & Yes & No & Yes \\ \cline{2-6}

& \textbf{Reference} & \cite{zhao2013} & \cite{zhao2019} & \cite{bala2019} &  \cite{zhang2023} \\ \hline \hline

\multirow{3}{*}{\textbf{Method}} & \textbf{Framework} & linear regression model with possible extension to generalized linear model  & linear regression model with possible extension to generalized linear model & linear regression model & linear, logistic and zero-inflated Poisson (ZIP) regression models \\ \cline{2-6}

& \textbf{Formulation} & $\mathcal{D}_i = \sum_{j=1}^{p} (\widehat{\rho}_j-\widehat{\rho}_j^{(i)})^2/p$,  where $\widehat{\rho}_j$ and $\widehat{\rho}_j^{(i)}$ are marginal correlation between $j^\text{th}$ predictor and response from the full and reduced dataset without $i^\text{th}$ observation. & modified HIM measure $\mathcal{D}_i$ designed to detect multiple outliers & $\delta_i = \sum_{j=1}^{p} \lvert \mathbbm{1}(\widehat{\beta}_{\text{LASSO},j}=0)$ - $\mathbbm{1}( \widehat{\beta}^{(i)}_{\text{LASSO},j}=0) \rvert$, where  $\widehat{\beta}_{\text{LASSO},j}$ and $\widehat{\beta}^{(i)}_{\text{LASSO},j}$ denote the $j^\text{th}$ component of the LASSO estimates from the full and reduced dataset without $i^\text{th}$ point. &  $\tau_i = \sum_{j=1}^{p} \lvert \mathbbm{1}(\widehat{\beta}_{j}=0)$ - $\mathbbm{1}( \widehat{\beta}^{(i)}_{j}=0) \rvert$, where $\widehat{\beta}_{j}$ and $\widehat{\beta}^{(i)}_{j}$ denote the $j^\text{th}$ component of a sparse estimate from the full and reduced dataset without $i^\text{th}$ observation. \\ \cline{2-6}

& \textbf{Key Step to Detect Multiple Outliers} & - & random-group-deletion (RDG) scheme (see Web Appendix B for details) & - & preliminary consistent high-dimensional clustering procedure \\ \cline{2-6}

& \textbf{Method Feature} & \makecell[l]{1. no model fitting \\ involved; \\ 2. simple implementation} & \makecell[l]{1. metrics based \\ on the HIM measure \\ shown to overcome \\ swamping and masking \\ ef{}fects; \\ 2. ef{}ficient RDG algorithm \\ to detect multiple outliers; \\ 3. no model fitting \\ involved} & \makecell[l]{1. explicitly captures a \\ single outlier on selection; \\ 2. framework applicable \\ to LASSO only; \\ 3. simple implementation} & \makecell[l]{1. explicitly captures \\ multiple outliers on \\ selection with generic \\ selectors; \\ 2. toolbox consisting of \\ ten parametric and non-\\parametric solutions to \\ approximate $\tau_i$; \\ 3. clustering facilitates \\ preliminary data division} \\ \hline \hline

\multirow{3}{*}{\textbf{Theory}} & \textbf{Key Result} & \makecell[l]{1.$n^2 \mathcal{D}_i$ is asymptotically \\ $\chi(1)$ as min$(n,p) \rightarrow \infty$; \\ 2. power of detection is \\ guaranteed under stated \\ assumptions} & \makecell[l]{1. asymptotic distribution \\ of modified HIM metrics \\ determined as min$(n,p)$ \\ $\rightarrow \infty$; \\ 2. FPR is controlled in \\ large-sample} & CLT holds for $\delta_i$ as  min$(n,p) \rightarrow \infty$ & \makecell[l]{1. CLT holds for $\tau_i$ \\ as min$(n,p) \rightarrow \infty$; \\ 2. $\tau_i$ is a mixture of \\ binomials asymptotically; \\ 3. FPR is controlled in \\ finite- and large-sample}  \\ \hline \hline

\multirow{2}{*}{\textbf{Computation}} & \textbf{Coding} & $\tt{R}$ package $\tt{ClusMIP}$ & $\tt{R}$ package $\tt{MIP}$ & $\tt{R}$ package $\tt{ClusMIP}$ & $\tt{R}$ package $\tt{ClusMIP}$  \\ \cline{2-6}

& \textbf{Cost} & low & low & low & moderate to high \\ \hline \hline

\multirow{3}{*}{\textbf{Performance}} & \textbf{Detection Capability} & (not included in the simulation studies) & satisfactory except when contamination exists in the predictors only or in case of highly-correlated predictors (even when signals are strong) & limited performance in the presence of multiple outliers under all scenarios included in the simulation studies & \makecell[l]{1. ClusMIP(SLASSO) \\ outperforms all other \\ procedures; \\ 2. ClusMIP applied to \\ SCAD and MCP exhibits \\ lower performance \\ depending on the \\ simulation setting} \\ \cline{2-6}

& \textbf{Strength} & theoretically justified approach for single outlier detection & ef{}fective and ef{}ficient in detecting multiple outliers when responses are contaminated & theoretically justified approach for single outlier detection & ef{}fective, theoretically justified and applicable to a wide range of model selectors \\ \cline{2-6}

& \textbf{Limitation} & not designed for detecting multiple outliers & limited detection capability when contamination exists in the predictors only & limitation in detecting multiple outliers under a wide range of contamination schemes & relatively high computational cost due to leave-one-out step in repeated model fitting \\ \hline \hline

\textbf{Utility} & \textbf{Usage} & when a single outlier exists and model-free approach is preferred & when contamination does not exist in the predictors and predictors are not highly-correlated & when a single outlier exists and the model selector is fixed to be LASSO & when having suf{}ficient computation resource and aiming to explore a range of outliers under dif{}ferent model selectors \\ \hline

\end{tabular}
\label{tab:method_comparison}
\end{table}

Consider the logistic regression model:
$\log\{\mathbb{P}(\text{Y}_i=1)/(1-\mathbb{P}(\text{Y}_i=1))\}=\mathbf{X}_i^\top\boldsymbol{\phi}$, $i=1,...,n$, 
where $\boldsymbol{\phi}=(\phi_1,\ldots,\phi_p)^\top$ and $\text{Y}_i\in\{0,1\}$. Outlier detection proceeds analogously via data partition and refined assessment. Unlike \eqref{linearmodel}, the mechanism under the logistic setting requires clustering tailored to the contamination type. The Expectation–Maximization algorithm may be appropriate when heterogeneity arises from finite mixtures of logistic regressions, though its sensitivity to initialization can hinder performance. Therefore, our simulations focus on contamination in predictors only, where the proposed clustering methods remain ef{}fective. 

\section{Simulation Study}
\label{sec:simulation}

%Here, we concisely describe the simulation setup and key patterns of the assessment metrics. 

We now perform extensive simulation studies to examine the performances of the two existing methods and the revised ClusMIP procedure under both linear and logistic regression models. The data generation details and all the graphical illustrations are given respectively in Web Appendices H.1-H.2 and H.4-H.6. Simulation settings are given in Web Tables S2 to S9. 

\subsection{Simulation Settings}
\label{sec:setting}

%n_{\text{inf\hspace{0.05em}l}}=
% $\text{Y}_{i}$, $i=1,\ldots,n$

\textcolor{black}{\textbf{Data}. The design matrix $\mathbf{X}=\mathbf{X}_{n,p}(\Sigma)$ is generated from a multivariate normal distribution with mean $\boldsymbol{0}$ and covariance $\Sigma$, where $\Sigma$ adopts the identity, autoregressive (AR), or equicorrelated (EQ) structures. The correlation parameters for AR and EQ covariances are chosen from $\{0.5,0.8\}$. Under \eqref{linearmodel}, $(n,p)\in\{(50,200),(100,1000),(120,1500)\}$, and under the logistic model, $(n,p)\in\{(100,200),(100,1000),(120,1500)\}$. Given $\mathbf{X}$ and the regression coef{}ficients, the responses are generated from the corresponding model. The first $10$ observations are then replaced with inf\hspace{+0.05em}luential points produced from Perturbation Schemes I–III, which integrate contamination into the responses, predictors, and both, with magnitudes from $\{5,10,30\}$. For the logistic regression model, only Scheme II is considered as discussed above}. 

%: the CLT, six parametric approaches (Param-CMB, Param-CMP, Param-BB, Param-GP, Param-MB, Param-MP), and three bootstrap schemes (Boot-I, Boot-II, Boot-III)

\noindent \textcolor{black}{\textbf{Method}}. \textcolor{black}{Under \eqref{linearmodel}, we have MIP, DF(LASSO), and the revised ClusMIP (with ten approximation options) applied to LASSO, SLASSO, SCAD and MCP. For the logistic regression, DF(LASSO) and the revised ClusMIP with LASSO, SCAD and MCP are considered. Performance is assessed by detection power, false positive rate, prediction and computation time, with prediction (before and after outlier removal) specifically measured by the mean squared error (MSE) for \eqref{linearmodel} and classification accuracy (ACC) for the logistic regression model}.

\subsection{Simulation Results}
\label{sec:simulation_results}

\begin{sidewaysfigure}
\includegraphics[width=1.05\textwidth]{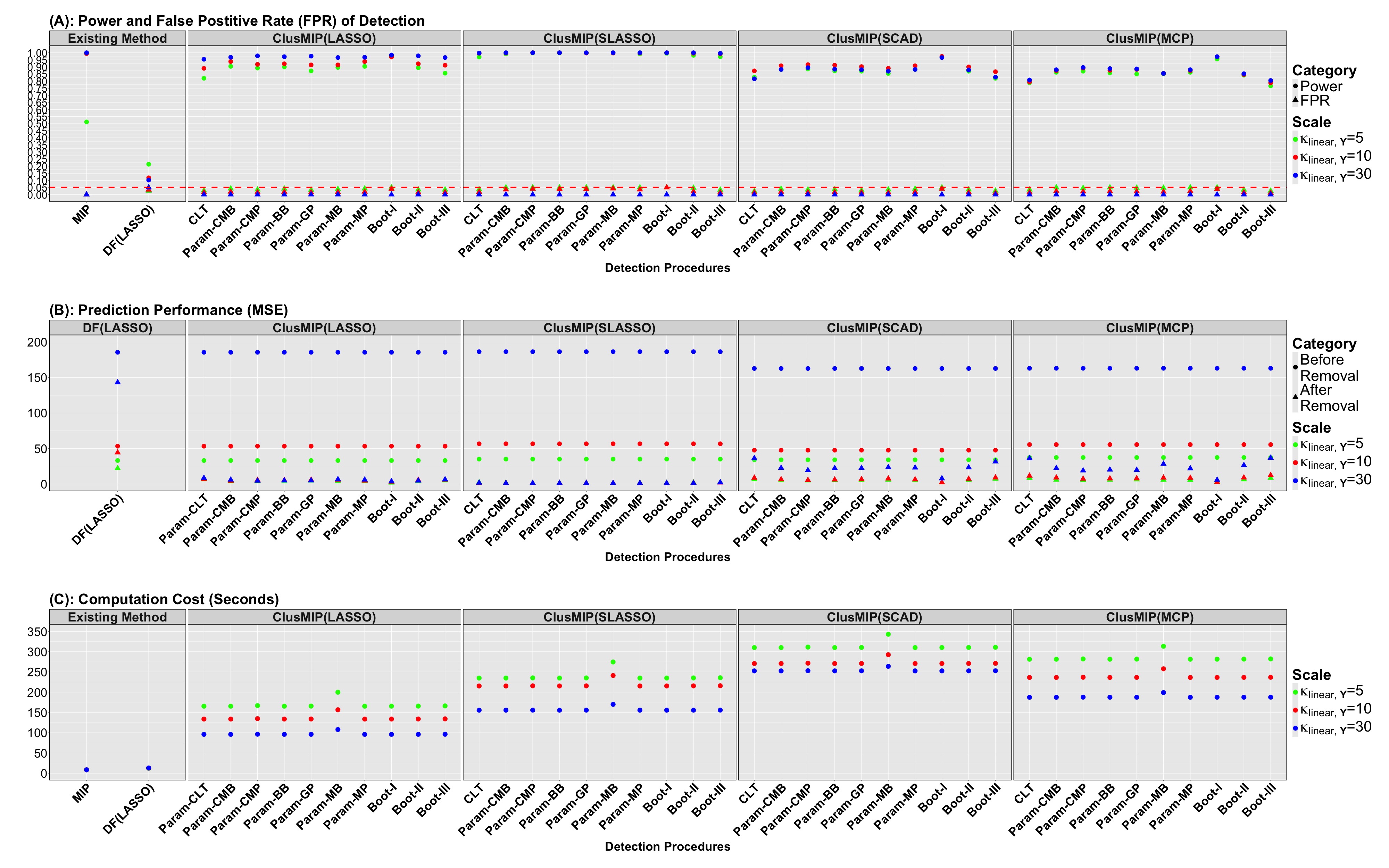} 
\caption{Linear regression model: simulation results (power and FPR) under $\textbf{X}(\Sigma(0.5))$ for Perturbation Models I (upper panel), II (middle panel) and III (lower panel). Here, ``Param-'' indicates the parametric approximation, while ``Boot-'' denotes the nonparametric approach.}
\label{fig:linear_I_AR8_n100_p1000}
\end{sidewaysfigure}

Under \eqref{linearmodel}, we present results for Perturbation Scheme I with $\mathbf{X}_{100,1000}(\Sigma)$, where $\Sigma$ follows an AR covariance with correlation 0.8. The illustration is shown in Figure~\ref{fig:linear_I_AR8_n100_p1000}. Unless otherwise specified, similar patterns are observed in other simulations, given in Web Appendix H.4-H.6.

\noindent In terms of detection power, DF(LASSO) consistently underperforms across simulations (below 0.25 as seen in Figure~\ref{fig:linear_I_AR8_n100_p1000}). In contrast, MIP performs well when contamination af{}fects responses, but underperforms when it is confined to predictors (Web Figure S25). In contrast, ClusMIP consistently yields stable and reasonable detection. Specifically, ClusMIP(SLASSO) almost always outperforms competing procedures across all perturbations and approximation schemes (almost 1.00 as observed in Figure~\ref{fig:linear_I_AR8_n100_p1000}). This is followed by ClusMIP(LASSO), which maintains power above 0.90 as seen in Figure~\ref{fig:linear_I_AR8_n100_p1000}. Overall, ClusMIP applied to convex penalties exhibits detection power comparable to or higher than that with concave penalties. Moreover, Boot-I, owing to its robustness, consistently attains the highest detection power among all the approximation options. Also, ClusMIP applied to all the selectors maintains false positive rates (FPRs) below the nominal level, substantiating Theorem 3 of \cite{zhang2023}. 

%, making them less susceptible to outliers
%Convex penalties impose uniform shrinkage, often leading to over-penalization of large signals and amplification of the impact of extreme residuals.

\textcolor{black}{The observed discrepancies in the detection performance of ClusMIP across dif{}ferent model selectors can be attributable to their varying sensitivities to outliers. Convex penalties impose uniform shrinkage, often over-penalizing large signals and amplifying the impact of extreme residuals. In particular, the joint estimation of coef{}ficients and noise variance may lead to outlier-induced variance inf\hspace{+0.05em}lation, further distorting selection for SLASSO. In contrast, concave penalties apply adaptive shrinkage that relaxes penalization for large coef{}ficients, preserving true signals and enhancing robustness. Moreover, while the selection consistency of LASSO requires the irrepresentable condition \citep{zhao2006} that can be easily violated by outlier-induced correlation distortion, concave penalties attain oracle properties under milder conditions \citep{fan2001,zhang2010}. Geometrically, convex penalties have rigid constraint boundaries such that outliers may push the optimization path toward extreme corners, whereas concave penalties have smoother contours and greater stability. Notably, the ClusMIP requires only data exchangeability, imposing minimal restrictions on the choice of model selectors. Yet, its detection ef{}fectiveness does rely on such choice, data characteristics, and the nature of contamination. See Web Appendix H.7 for detailed discussions}. 

\textcolor{black}{Clear reductions in MSE are observed after removing detected inf\hspace{+0.05em}luential points, as the refit models are no longer ``pulled'' toward outliers, thereby improving fit for uncontaminated data. In Figure~\ref{fig:linear_I_AR8_n100_p1000}, the mean MSEs decrease by $22.9\%$, $97.1\%$, $99.4\%$, $85.8\%$, and $85.6\%$ for DF(LASSO) and ClusMIP applied to LASSO, SLASSO, SCAD, and MCP, respectively. The extent of MSE reduction corresponds closely to the detection power, indicating that removing more true inf\hspace{+0.05em}luential observations enhances model generalizability. At perturbation level 30, detection powers of DF(LASSO) and ClusMIP with LASSO, SLASSO, SCAD, and MCP are $0.102$, $0.971$, $0.999$, $0.878$, and $0.872$ respectively, consistent with the observed MSE declines}.

\textcolor{black}{In terms of computation, ClusMIP is substantially more time-consuming than the existing methods. At perturbation level 30 (Figure~\ref{fig:linear_I_AR8_n100_p1000}), the mean runtimes for MIP, DF(LASSO), and ClusMIP with LASSO, SLASSO, SCAD, and MCP are 8.4, 13.0, 97.2, 157.0, 253.8, and 188.6 seconds, respectively. Overall, ClusMIP(SCAD) has the longest runtime (1161.80 seconds in Figure S49), while the existing methods remain below 21 seconds. The discrepancy is primarily due to repeated leave-one-out model fitting required by ClusMIP. Also, ClusMIP applied to concave penalties is slower than convex penalties, as concave optimization may involve multiple local extrema, requiring iterative or multistart strategies for stable convergence}.

Under the logistic regression model, the observed patterns of the four metrics are analogous to those under \eqref{linearmodel} explained above. Their graphical results are given in Web Appendix H.5. 

\section{Pain Prediction using fMRI Data}
\label{sec:real}

We now apply the proposed method to an fMRI dataset examining brain responses to physical pain, with the aim of detecting outliers and evaluating their impact on downstream analyses. 

\noindent \emph{\bf Description}. The experiment consisted of repeated thermal stimulation at six temperatures ($44.3, 45.3, 46.3, 47.3, 48.3$, and $49.3 \ \tccentigrade$) given to the left forearm of 33 participants. After each stimulus, self-reported pain was recorded on a 200-point scale, where values above and below 100 indicate high and low pain ratings, respectively. Pain scores were treated as our outcome of interest. Technical specifics on fMRI data acquisition, preprocessing, and analysis are given in \cite{lindquist2017}. In short, within-subject voxel-wise functional activation maps were estimated for each stimulus. The maps were then averaged across trials at each temperature, yielding six maps per subject, and further spatially averaged across 489 brain regions. This gives a design matrix with 489 predictors, corresponding to the average temperature-specific brain activity in a specific brain region for each participant, and 198 pain score responses.  

\noindent \emph{\bf Method}. The dataset is analyzed by both linear and logistic regression models. For the linear model, we include DF(LASSO), MIP and the revised ClusMIP (containing ten approximation options) applied to LASSO, SLASSO, SCAD and MCP. For the logistic regression model, binary responses are defined by coding pain ratings below 100 as 0 and above 100 as 1, and DF(LASSO) and the revised ClusMIP applied to LASSO, SCAD and MCP are considered.  

\noindent \textcolor{black}{\emph{\bf Assumption}}. \textcolor{black}{This dataset exhibits complex within- and between-subject correlations. Yet, several lines of evidence, including the presence of randomized and counterbalanced stimulus presentation, suf{}ficient rest time, well-controlled experimental conditions, limited systematic bias in pain experience attributable to sex, race, or genetics, and a shared biological pathway of pain perception, collectively support treating exchangeability as a tentative working assumption that reasonably accommodates correlations. Consistent with the core assumption of ClusMIP, exchangeability is plausible when measurements are collected under comparable conditions, including repeated-measure designs with standardized protocols, harmonized multi-center cohort studies, or imaging experiments, where variability is primarily random. Penalized linear regression models remain practically useful under exchangeability, and the leave-one-subject-out scheme further addresses dependence to some extent. Thus, despite the existence of sophisticated methods designed to handle within-subject correlation, our method of{}fers a plausible starting point for modeling pain, especially for outlier diagnostics, consistent with prior studies \citep{geuter2020}. Detailed discussion is given in Web Appendix I.1}. 

\begin{figure}[H]
\noindent\makebox[\textwidth]{%
	%\begin{center}
		\includegraphics[width=1.05\textwidth]{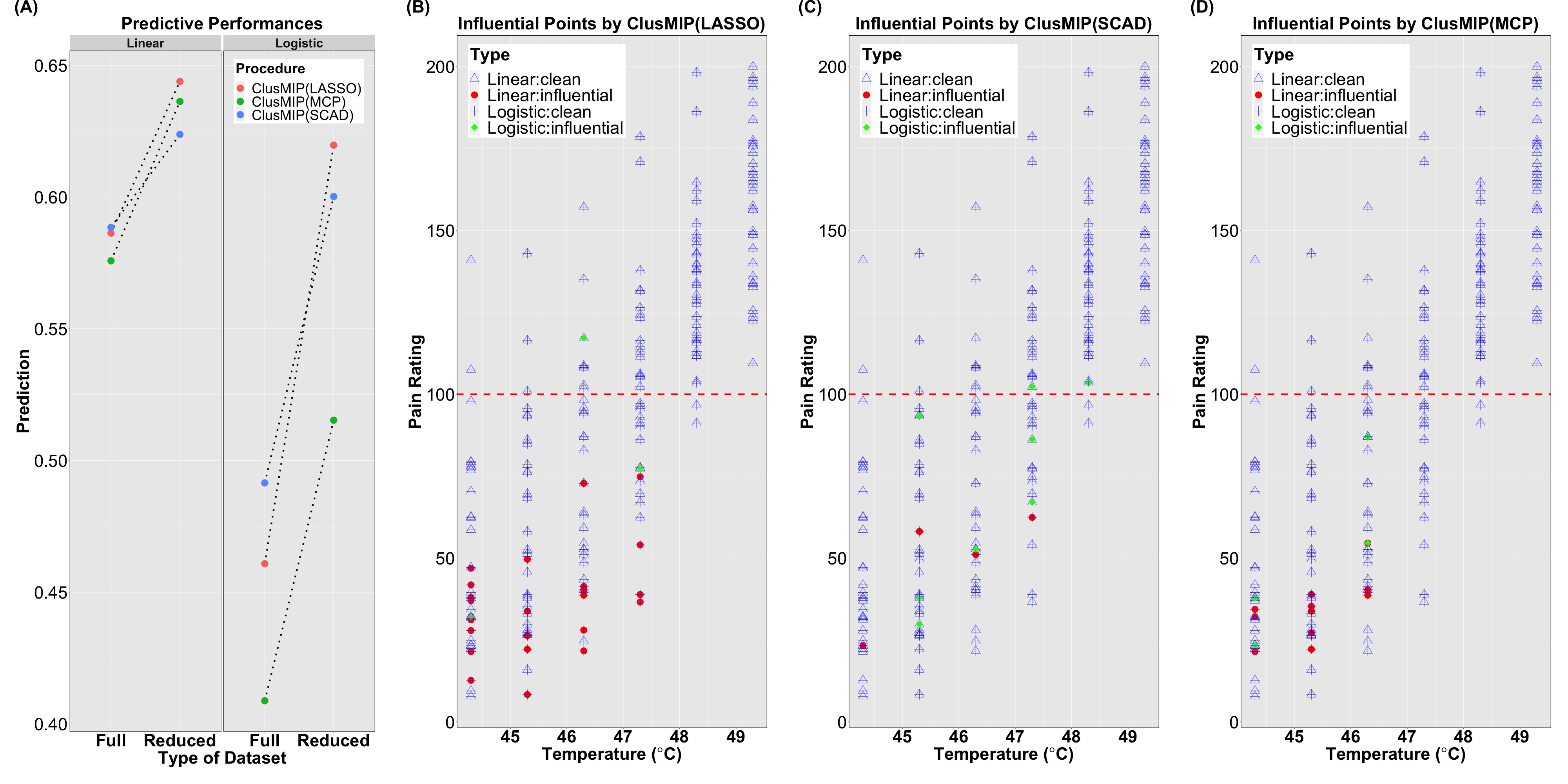}}
	%\end{center}}
	\caption{Real data analysis on pain prediction using fMRI data. Panel (A): comparison of predictive performance before and after removing detected inf\hspace{+0.05em}luential points under linear and logistic regression models. Panels (B) to (D): for the revised ClusMIP applied to the LASSO, SCAD and MCP, the distribution of detected inf\hspace{+0.05em}luential observations and the remaining clean observations under linear and logistic regression models mapped back to the pain ratings classified according to the six temperatures.}
	\label{fig:real_data_analysis_combined_plot}
\end{figure}

\begin{figure}[H]
\noindent\makebox[\textwidth]{%
	%\begin{center}
		\includegraphics[width=1.05\textwidth]{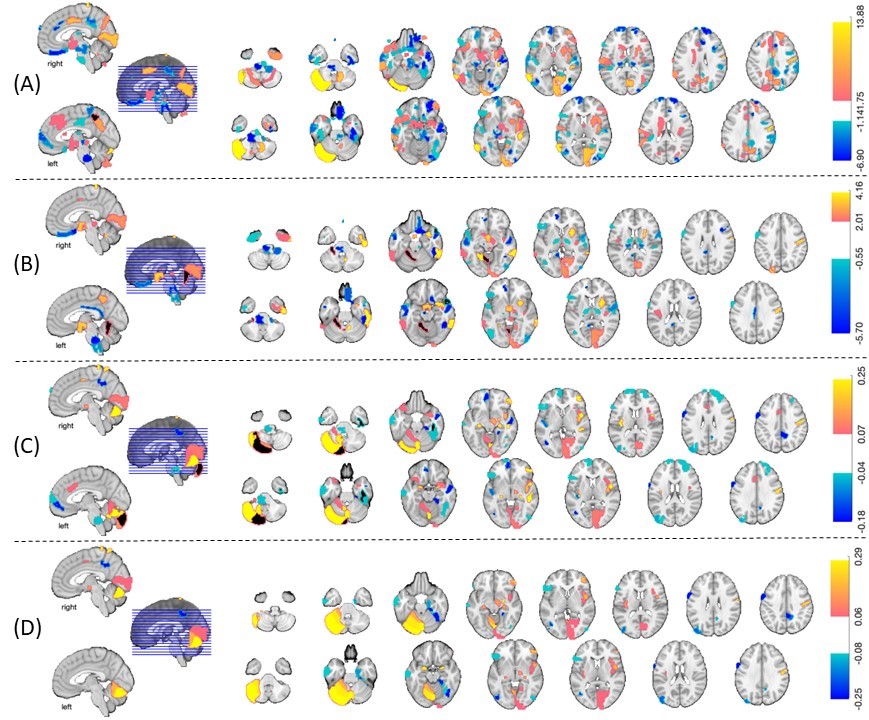}}
	%\end{center}}
	\caption{Real data analysis on pain prediction using fMRI data: selection of brain regions and the corresponding magnitude of the LASSO regression coefficients. Panel (A): linear regression model based on the full dataset. Panel (B): linear regression model based on the reduced dataset upon removing inf\hspace{+0.05em}luential points given by $\widehat{\text{I}}_{\text{linear}}(\text{ClusMIP(LASSO)})$ with Boot-I. Panel (C): logistic regression model based on the full dataset. Panel (D): logistic regression model based on the reduced dataset upon removing inf\hspace{+0.05em}luential points given by $\widehat{\text{I}}_{\text{logistic}}(\text{ClusMIP(LASSO)})$ with Boot-I. Here, Boot-I is the first bootstrap scheme to derive threshold of $\tau_{[n]}$ for diagnosis purposes discussed in Section~\ref{sec:nonparametric}.}
	\label{fig:real_data_analysis_brain_selection}
\end{figure}

\noindent \emph{\bf Assessment}. We tabulate the estimated inf\hspace{+0.05em}luential points denoted by $\widehat{\text{I}}_{\text{linear}}(\text{Detection})$ and $\widehat{\text{I}}_{\text{logistic}}(\text{Detection})$ for both models in Web Tables~S10 to S13, and compute predictive metrics (Figure~\ref{fig:real_data_analysis_combined_plot}) and contrast the selected brain regions before and after their exclusion (Figure~\ref{fig:real_data_analysis_brain_selection}. 

%following $\lvert \widehat{\text{I}}_{\text{linear}}(\text{ClusMIP(LASSO)}) \rvert > \\ \lvert \widehat{\text{I}}_{\text{linear}}(\text{ClusMIP(SCAD)}) \rvert \approx \lvert \widehat{\text{I}}_{\text{linear}}(\text{ClusMIP(MCP)}) \rvert$
% (except under Param-CMP, GP, MP, and Boot-I, which detect few)
\noindent \emph{\bf Detection}. Considerable variation in the identified outliers exists across all procedures. Few outliers are detected by MIP or ClusMIP(SLASSO), consistent with simulations that they exhibit similar performance when responses are contaminated. In contrast, ClusMIP applied to LASSO, SCAD, and MCP captures substantially more outliers, where $\widehat{\text{I}}_{\text{linear}}(\text{ClusMIP(LASSO)})$ is the largest. Among the ten approximation options, Boot-I consistently identifies the largest number of outliers, corresponding to 76, 8, 53, and 59 inf\hspace{+0.05em}luential points for ClusMIP applied to LASSO, SLASSO, SCAD and MCP, respectively. This aligns with its superior detection power observed in simulations. Yet, under the logistic regression model, substantially fewer outliers are identified, with Boot-I still attaining the highest numbers of detected outliers. 

\noindent \emph{\bf Distribution}. Under both linear and logistic regression models, for each selector, we identify common outliers detected by ClusMIP across the ten approximation options (Table~\ref{tab:real_data_detection}). They are shown in the pain-rating plot at six temperatures in Panels (B) to (D) of Figure~\ref{fig:real_data_analysis_combined_plot}, where we see that outlying points primarily occur at lower pain levels ($<100$) corresponding to lower temperatures ($44.3, 45.3$ and $46.3 0\ \tccentigrade$), consistent with the concept that weak pain responses to warm stimuli may follow a dif{}ferent pattern than those at higher temperatures.

\begin{table}[H]
\centering
 \addtolength{\leftskip}{-2cm}
    \addtolength{\rightskip}{-2cm}
    \fontsize{9}{11}\selectfont
    \caption{\centering Indices of Estimated Influential Observations for the Physical Pain Prediction Dataset. \\ For the ClusMIP procedure, they are obtained via extracting the commonly identified influential points among the ten approximating approaches discussed in Sections~\ref{sec:parametric} and \ref{sec:nonparametric}.}
\begin{tabular}{C{2.8cm} C{6cm} C{4.5cm}}
  \hline
  \textbf{Procedures} & Common $\widehat{\text{I}}_{\text{linear}}$ & Common $\widehat{\text{I}}_{\text{logistic}}$ \\ 
  \hline
 % \hline
    DF(LASSO) & 2,6,72,121,130,138,174,181 & 23,73 \\ 
    \hline
    MIP & NA & -  \\ 
    \hline  
    (Revised) ClusMIP(LASSO) & \makecell{7,13,14,43,45,50,51,52,55,56,57,58,82,87,\\103,104,127,133,134,145,147,171,172,175} & 13,111,166  \\ 
    \hline
    (Revised) ClusMIP(SLASSO) & NA & -  \\ 
    \hline
    (Revised) ClusMIP(SCAD) & 44,124,163,165 & 4,22,74,83,135,148,170,188  \\ 
    \hline
    (Revised) ClusMIP(MCP) & 14,15,19,20,56,57,87,115,145,146,176 & 3,15,73,157  \\ 
   \hline
\end{tabular}
\label{tab:real_data_detection}
\end{table}

\noindent \emph{\bf Prediction}. Under both linear and logistic models, we assess the prediction performance before and after removal of detected outliers shown in Table~\ref{tab:real_data_detection}. Specifically, both the full and reduced dataset are randomly split into  training and testing sets: regression coef{}ficients are obtained on the training set while the predictive metrics are calculated on the test set. Under the linear model, prediction is measured by the Pearson's correlation between observed and predicted pain scores; under the logistic model, it is evaluated by the classification accuracy of binary pain responses. Each metric is averaged over 1000 random splits, yielding the results in Panel (A) of Figure~\ref{fig:real_data_analysis_combined_plot}. This figure shows at least a $10\%$ improvement in the predictive performance after outlier removal for both models, underscoring the value of their detection.   

\noindent \emph{\bf Variation in selected brain regions}. Under both linear and logistic regression models, discrepancy in the selected brain regions and magnitudes of regression coef{}ficients is observed before and after removing $\widehat{\text{I}}_{\text{linear}}$ and $\widehat{\text{I}}_{\text{logistic}}$. While such variation exists across all detection methods, it is illustrated based on ClusMIP(LASSO) with Boot-I shown in Figure~\ref{fig:real_data_analysis_brain_selection}. 
The maps from the reduced dataset show improved quality. Specifically, fewer regions near brain boundary are selected that are typically driven by excessive motion artifacts, while key pain-processing regions (e.g., somatosensory S1/S2, medial thalamus, anterior cingulate, and mid-insular-opercular areas) are retained. These findings remain consistent under both models.

%. This can be seen by the reduction of selected regions near the boundary of the brain, which are typically activated by excessive subject motion, while retaining the key regions know to be involved in pain processing pathways (e.g., somatosensory S1/S2, medial thalamus, Anterior Cingulate, and mid insular-opercular areas). These findings are consistent for both models.

%Real data analysis on pain prediction using fMRI data: selection of brain regions and the corresponding magnitude of the LASSO regression coefficients. Panel (A): linear regression model based on the full dataset. Panel (B): linear regression model based on the reduced dataset upon removing influential points given by $\widehat{\text{I}}_{\text{linear}}(\text{ClusMIP(LASSO)})$ with Boot-I. Panel (C): logistic regression model based on the full dataset. Panel (D): logistic regression model based on the reduced dataset upon removing influential points given by $\widehat{\text{I}}_{\text{logistic}}(\text{ClusMIP(LASSO)})$ with Boot-I. Here, Boot-I is the first bootstrap scheme to derive threshold of $\tau_{[n]}$ for diagnosis purposes discussed in Section~\ref{sec:nonparametric}.

%stemming from its uniform shrinkage and stringent requirement on selection consistency, 

\noindent \emph{\bf \textcolor{black}{Guidelines on model selector choice}}. \textcolor{black}{In general, the choice of model selector depends on data characteristics and modeling objectives. In analyzing the thermal pain dataset, the primary objective is to capture inf\hspace{+0.05em}luential observations distorting variable selection. To this end, LASSO is a preferable candidate due to its heightened sensitivity to outliers as discussed in Section~\ref{sec:simulation_results}. Also, removing outliers detected by the revised ClusMIP applied to LASSO yields the best predictive performance (Figure~\ref{fig:real_data_analysis_combined_plot}). The corresponding brain regions selected after excluding detected outliers (Figure~\ref{fig:real_data_analysis_brain_selection}) are also scientifically plausible. As such, LASSO appears to be the most favorable candidate when combined with the revised ClusMIP}.

%Variation in the model-dependent detection of influential points necessitates guidance on which outliers to consider.
%based on exploratory analysis and domain knowledge

\noindent \emph{\bf Guidelines on outlier choice}. In practice, practitioners may choose a toolbox of model selectors \emph{a priori}. Based on these model selectors, our framework enables flexible aggregation of detected outlier sets, ranging from intersection to union of these collections, depending on the desired conservativeness. In the thermal pain study, although LASSO is the recommended selector, our framework indeed produces multiple outlier sets, with those commonly identified deemed most important. Under Boot-I, further intersection of $\widehat{\text{I}}_{\text{common}} \coloneq \widehat{\text{I}}_{\text{linear}} \cap \widehat{\text{I}}_{\text{logistic}}$ across ClusMIP applied to LASSO, SCAD and MCP yields $\{49,55,61,169\}$. Compared with \cite{zhang2023}, two additional inf\hspace{+0.05em}luential observations $\{61,169\}$ are captured which corresponds to $44.3 \ \tccentigrade$. This reinforces that outliers often occur at lower temperatures where not all participants report pain. The availability of a broad spectrum of outliers through ClusMIP reflects our philosophy as statistical toolbox providers rather than decision makers, consistent with Fisher's and Tukey’s view that we ``of{}fer guidance, not the answer'' \citep{tukey1962}. 

\section{Conclusions}
\label{sec:conclusions}

%Application to thermal pain study data further strengthens previously established findings.

This work extends \cite{zhang2023} to capture inf\hspace{+0.05em}luential observations af{}fecting variable selection. Specifically, the asymptotic distribution of the DF(LASSO) and GDF \eqref{gdf} measures are established in Theorem~\ref{thm:tau_asymptotic}. Six parametric (supported by Theorem~\ref{thm:kldistance}) and three bootstrap schemes are proposed to approximate the distribution of $\tau_i$, from which diagnostic thresholds in the form of mid-quantile are derived, resolving the inconsistency induced by discreteness. Integrating these approximations into the ClusMIP algorithm leads to an updated version, implemented in the $\tt{R}$ package $\tt{ClusMIP}$. Simulation results indicate strong detection ef{}fectiveness under linear and logistic regression models. \textcolor{black}{Notably, our approach also performs reasonably well under the zero-inf\hspace{+0.05em}lated Poisson regression model (Web Appendix H.6)}. Based on this work, the following insights are obtained. 

\noindent 1. {\bf \textcolor{black}{Feature of the GDF Measure}}. \textcolor{black}{The distributional properties of $\tau_i$ depend solely on data exchangeability and are invariant to factors such as collinearity. Yet, the diagnostic ef{}fectiveness of $\tau_i$ does depend in subtle ways on data characteristics, model selector choice, signal strength, and the nature of contamination. See Web Appendix J for detailed discussion}. 

%than those of ClusMIP applied to LASSO and SLASSO

\noindent 2. {\bf Detection Methods}. DF(LASSO) underperforms due to its leave-one-out design, whereas MIP performs well except when contamination af{}fects only the predictors. Among the ClusMIP variants, ClusMIP(SLASSO) consistently achieves strong detection performance due to uniform shrinkage and stricter requirements on selection consistency, followed by ClusMIP(LASSO). Thus, they are suggested for detection purposes. In contrast, ClusMIP applied to SCAD or MCP has lower detection power than that of ClusMIP applied to convex penalties, owing to their adaptive shrinkage and relaxed requirements on selection consistency. % (oracle properties). 

%\noindent 3. {\bf ClusMIP Procedure}. Due to uniform shrinkage and stricter requirements on selection consistency, ClusMIP(SLASSO) attains consistent satisfactory performance, followed by ClusMIP(LASSO). Therefore, they are recommended for detection purposes. In comparison, detection power of ClusMIP applied to SCAD or MCP is lower than those of ClusMIP applied to LASSO and SLASSO owing to their adaptive shrinkage and relaxed requirements on model selection consistency (oracle properties). 

%Simulation results under both linear and logistic models show that Boot-I consistently achieves the highest detection power, establishing it as the preferred default approximation

\noindent 3. {\bf Approximation Option}. The approximation choice relies on dimensionality constraints, interpretability, robustness and computational cost. While the bootstrap schemes are robust and impose minimal dimensionality restrictions (requiring only large $n$), the parametric options improve interpretability by characterizing the latent structure of the diagnostic measures. Simulation results under both linear and logistic models show that Boot-I consistently achieves the highest detection power, establishing it as the preferred default approximation.

%indicate Boot-I as the preferred default choice, owing to the highest detection power observed. 
%consistently has the highest detection power, establishing it as the preferred default approximation. 
%under both linear and logistic regression models

%Yet, variations in the detection outcomes of the revised ClusMIP across dif{}ferent selectors remain to be theoretically studied

The existing and proposed methods are further applied to a dataset from an fMRI-based pain study \citep{lindquist2017}. Consistent with the simulation results, predictive performances are improved after removing the identified outliers. Yet, variations in the detection outcomes of the revised ClusMIP across dif{}ferent selectors remain to be theoretically studied.

%  If your paper refers to supporting web material, then you MUST
%  include this section!!  See Instructions for Authors at the journal
%  website http://www.biometrics.tibs.org

\section*{Acknowledgments}

The authors thank the Editor, the Associate Editor and two anonymous referees for valuable comments and suggestions. 

\section*{Supplementary Materials}

Web Appendices and code referenced in Sections 1-6 are available with this paper at the Biometrics website on Oxford Academic. An $\tt{R}$ package  $\tt{ClusMIP}$ containing implementation for the revised ClusMIP procedure is given on GitHub: https://github.com/Dongliang-JHU/ClusMIP. 

\section*{Funding}

This work was supported, in part, by the National Institute of Mental Health (R01MH129397), the National Institute of Biomedical Imaging and Bioengineering (R01EB026549), and the Natural Science and Engineering Research Council of Canada (NSERC RGPIN-2018-05618).

%Dongliang Zhang is partially supported by NIH grant R01MH129397 from the National Institute of Mental Health. Martin A.~Lindquist is supported in part by NIH grant R01 EB026549 from the National Institute of Biomedical Imaging and Bioengineering and R01MH129397 from the National Institute of Mental Health. Masoud Asgharian is supported by the Natural Science and Engineering Research Council of Canada (NSERC RGPIN-2018-05618). \vspace*{-8pt}

\section*{Data Availability}

The data that support the findings in this paper will be shared on reasonable request to the corresponding author.

\bibliography{References}
	
\end{document}